\documentclass[journal=jacsat,manuscript=article]{achemso}

\usepackage{chemformula} % Formula subscripts using \ch{}
\usepackage[T1]{fontenc} % Use modern font encodings
\usepackage{braket}
\usepackage{lineno}
\usepackage[most]{tcolorbox}
\usepackage{subfig}

\usepackage{amssymb}

\author{Sumukh Vaidya}
\author{Xingyu Gao}
\affiliation{Department of Physics and Astronomy, Purdue University, West Lafayette, Indiana 47907, USA}
\author{Saakshi Dikshit}
\affiliation{Elmore Family School of Electrical and Computer Engineering, Purdue University, West Lafayette, Indiana 47907, USA}
\author{Zhenyao Fang}
\affiliation{Department of Physics, Northeastern University, Boston,
Massachusetts 02115, USA}

\author{Andres E Llacsahuanga Alcca}
\affiliation{Department of Physics and Astronomy, Purdue University, West Lafayette, Indiana 47907, USA}
\author{Yong P Chen}
\affiliation{Department of Physics and Astronomy, Purdue University, West Lafayette, Indiana 47907, USA}
\alsoaffiliation{Elmore Family School of Electrical and Computer Engineering, Purdue University, West Lafayette, Indiana 47907, USA}
\alsoaffiliation{Purdue Quantum Science and Engineering Institute,
Purdue University, West Lafayette, Indiana 47907, USA}
\alsoaffiliation{Birck Nanotechnology Center, Purdue University, West Lafayette, Indiana 47907, USA}
\author{Qimin Yan}
\affiliation{Department of Physics, Northeastern University, Boston,
Massachusetts 02115, USA}

\author{Tongcang Li}
\email{tcli@purdue.edu}
\affiliation{Department of Physics and Astronomy, Purdue University, West Lafayette, Indiana 47907, USA}
\alsoaffiliation{Elmore Family School of Electrical and Computer Engineering, Purdue University, West Lafayette, Indiana 47907, USA}
\alsoaffiliation{Purdue Quantum Science and Engineering Institute,
Purdue University, West Lafayette, Indiana 47907, USA}
\alsoaffiliation{Birck Nanotechnology Center, Purdue University, West Lafayette, Indiana 47907, USA}

\title[2025 $GeS_2$ paper]
    {Coherent Spins in van der Waals Semiconductor GeS$_2$ at Ambient Conditions}

\abbreviations{IR,NMR,UV}

\begin{document}

%%%%%%%%%%%%%%%%%%%%%%%%%%%%%%%%%%%%%%%%%%%%%%%%%%%%%%%%%%%%%%%%%%%%%
%% The "tocentry" environment can be used to create an entry for the
%% graphical table of contents. It is given here as some journals
%% require that it is printed as part of the abstract page. It will
%% be automatically moved as appropriate.
%%%%%%%%%%%%%%%%%%%%%%%%%%%%%%%%%%%%%%%%%%%%%%%%%%%%%%%%%%%%%%%%%%%%%
% \begin{tocentry}

% Some journals require a graphical entry for the Table of Contents.
% This should be laid out ``print ready'' so that the sizing of the
% text is correct.

% Inside the \texttt{tocentry} environment, the font used is Helvetica
% 8\,pt, as required by \emph{Journal of the American Chemical
% Society}.

% The surrounding frame is 9\,cm by 3.5\,cm, which is the maximum
% permitted for  \emph{Journal of the American Chemical Society}
% graphical table of content entries. The box will not resize if the
% content is too big: instead it will overflow the edge of the box.

% This box and the associated title will always be printed on a
% separate page at the end of the document.

% \end{tocentry}

%%%%%%%%%%%%%%%%%%%%%%%%%%%%%%%%%%%%%%%%%%%%%%%%%%%%%%%%%%%%%%%%%%%%%
%% The abstract environment will automatically gobble the contents
%% if an abstract is not used by the target journal.
%%%%%%%%%%%%%%%%%%%%%%%%%%%%%%%%%%%%%%%%%%%%%%%%%%%%%%%%%%%%%%%%%%%%%
\begin{abstract}
  Optically active spin defects in van der Waals (vdW) materials have recently emerged as versatile quantum sensors, enabling applications from nanoscale magnetic field detection to the exploration of novel quantum phenomena in condensed matter systems. Their ease of exfoliation and compatibility with device integration make them promising candidates for future quantum technologies. Here we report the observation and room-temperature coherent control of spin defects in the high-temperature crystalline phase of germanium disulfide ($\beta$-GeS$_2$), a two-dimensional (2D) semiconductor with low nuclear spin density. The observed spin defects exhibit spin-1/2 behavior, and their spin dynamics can be explained by a weakly coupled spin-pair model. We implement dynamical decoupling techniques to extend the spin coherence time (T$_2$) by a factor of 20.  Finally, we use density functional theory (DFT) calculations to estimate the structure and spin densities of two possible spin defect candidates. This work will help expand the field of quantum sensing with spin defects in van der Waals materials. 
\end{abstract}

%%%%%%%%%%%%%%%%%%%%%%%%%%%%%%%%%%%%%%%%%%%%%%%%%%%%%%%%%%%%%%%%%%%%%
%% Start the main part of the manuscript here.
%%%%%%%%%%%%%%%%%%%%%%%%%%%%%%%%%%%%%%%%%%%%%%%%%%%%%%%%%%%%%%%%%%%%%

%%%%%%%%%%%%%% 
% Introduction
\section{Table of Contents (TOC) graphic}
% \red{Some Basic intro stuff...., $GeS_2$ predicted $T_2$ coherence time is 4.51 ms \cite{kanai2022generalized, sajid2022spin}}
\includegraphics[scale = .2]{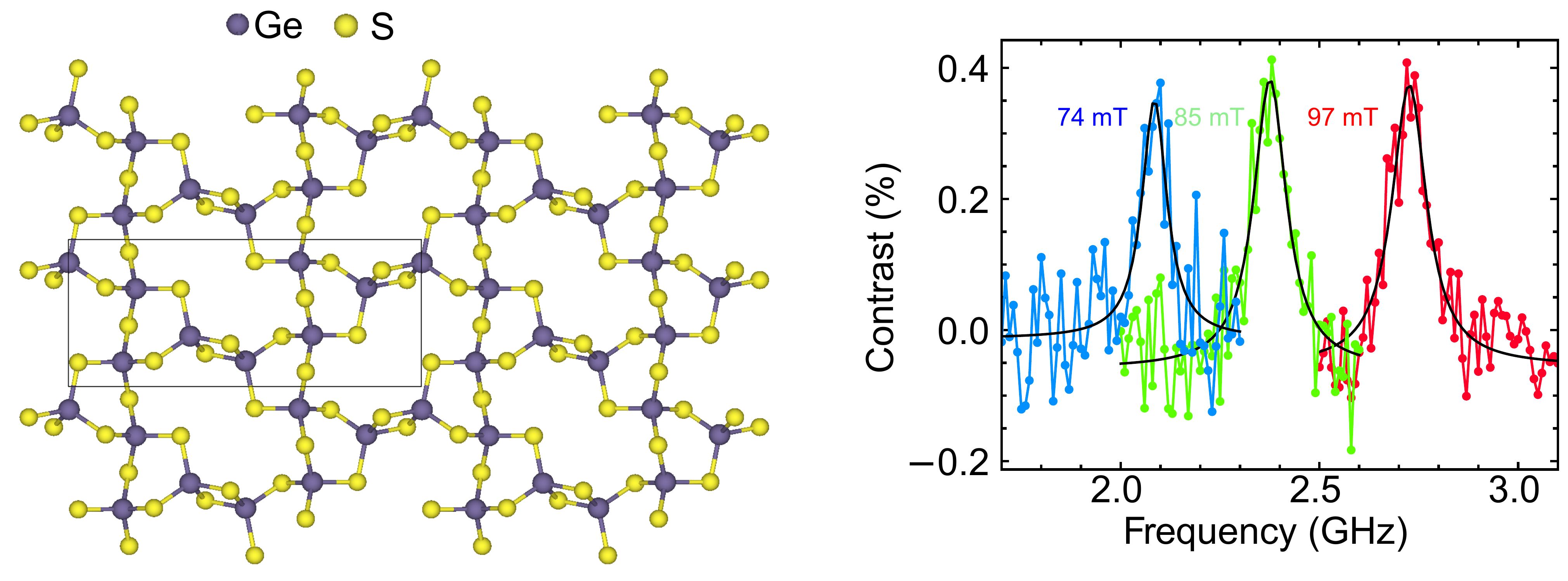}

\section{Main text}
Optically active solid-state spin-defects are a frontier in quantum technology. Owing to efficient spin-photon interface and scalability in terms of crystal growth and device integration, they have seen extensive interest in the past few years. Some prominent examples are the negatively charged nitrogen vacancy (NV$^-$) center in diamond \cite{schirhagl2014nitrogen}, the silicon vacancy (V$_{Si}$)  in silicon carbide (SiC) \cite{wang2023magnetic}, and more recently the T-center in silicon \cite{higginbottom2022optical}. These spin defects have been used to demonstrate a wide variety of applications in quantum sensing \cite{schirhagl2014nitrogen, wang2023magnetic}, networking\cite{togan2010quantum, stas2022robust,   higginbottom2022optical}, and simulation\cite{cai2013large, rong2015experimental, wu2019observation}. After the discovery of the first optically active spin defects in van der Waals (vdW) material hexagonal boron nitride (hBN) \cite{gottscholl2020initialization, chejanovsky2021single, gao2021high, stern2022room}, namely the negatively charged boron vacancy (V$_B^-$), the field of two dimensional (2D) materials-based spin defect color centers has also seen a lot of interest. hBN has also been found to host several other spin defects apart from V$_B^-$ center \cite{mendelson2021identifying,chejanovsky2021single, liu2022spin,stern2022room, gao2024single, guo2023coherent,  stern2024quantum, scholten2024multi} . 2D quantum sensing with hBN has been used to study a wide range of interesting physical phenomena including, but not limited to magnetic fields, strain, temperature, nuclear spins and RF signals \cite{gottscholl2021spin, gao2022nuclear, huang2022wide, kumar2022magnetic, healey2023quantum, robertson2023detection, zhou2023dc, gao2023quantum, vaidya2023quantum}. Pushing further towards the 2D limit, few-layer hBN has also been demonstrated to host sensing-suitable spin qubits, which makes it ideal for proximal sensing of novel physical phenomena \cite{zhou2024sensing,durand2023optically}. 

\begin{figure}
  \includegraphics[scale = .55]{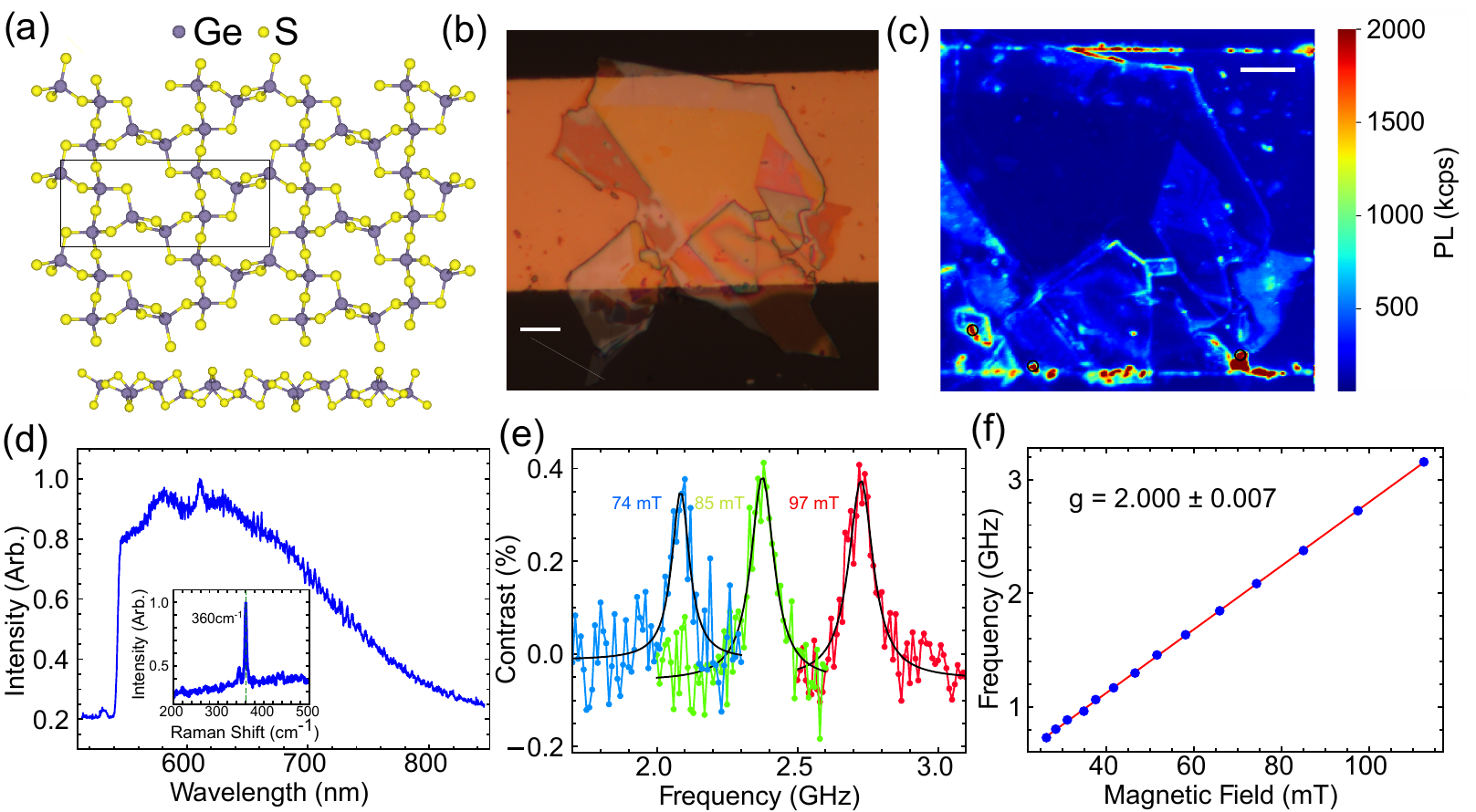}
  \caption{(a) The crystal structure of $\beta$-GeS$_2$. The black box in the top view denotes the unit cell and the lower figure denotes the side view of the lattice. (b) Microscope Images of the $\beta$-GeS$_2$ sample studied in the experiment. The scale bar is 10 $\mu$m. (c) Confocal scans of the sample under 532 nm excitation. The scale bar is 10 $\mu$m. (d) Photoluminescence spectrum of the spin defects under 532 nm laser excitation. The inset shows the Raman spectrum with a sharp peak at 360 cm$^{-1}$. (e) ODMR spectrum measured under 532 nm laser and 1 W of MW excitation. The black curves are fits of the data to a Lorentzian function. (f) Dependence of the ODMR peak frequency on an external magnetic field  perpendicular to the sample surface. }
  \label{figure_1}
\end{figure}

% In order to realize these applications of spin color centers, it is imperative to utilize long spin coherence times. 
hBN quantum sensors are however currently limited in their scope as highly useful qubits on account of their short coherence times \cite{gottscholl2021room, gao2021high, haykal2022decoherence, vaidya2023quantum}.
The primary source of decoherence in solid state spins is the nuclear bath that surrounds the electron spin. The V$_B^-$ spin defects in hBN for instance have a spin relaxation time ($T_1$) of 18 $\mu$s and a spin coherence time ($T_2$) of approximately 100 ns \cite{gottscholl2021room, haykal2022decoherence}, on account of the boron and nitrogen nuclei that form the crystal lattice. This in turn has inspired the search for optically active spin defects in other 2D materials as well, with the aim of finding spin defects with long coherence times \cite{kanai2022generalized, sajid2022spin}. The versatility of 2D materials based quantum sensors in terms of proximity to a sample under measurement and from a process integration perspective makes it highly desirable to search for spin qubits in other 2D material platforms. The high-temperature phase $\beta$-germanium disulfide ($\beta$-GeS$_2$) is a 2D semiconductor with a bandgap of 3.2-3.7 eV\cite{mitsa2014spectroscopic, yang2019polarization, slavich2025germanium}. The most abundant isotopes of germanium and sulfur have nuclear spin 0, suggesting potentially lower spin noise and longer resultant coherence times. Cluster Correlation Expansion (CCE) calculations indicate that coherence time $T_2$ of spin defects in $\beta$-GeS$_2$ could be as high as 4.5~ms \cite{kanai2022generalized, sajid2022spin, fabian2022quantum}, which would make it a highly ideal material platform for both quantum computing and sensing applications. 

In this paper we report the observation and coherent control of spin defects in the vdW material $\beta$-GeS$_2$ at room temperature. We use exfoliation and high-temperature annealing to prepare the samples and transfer the samples onto a gold microwave stripline antenna for efficient microwave delivery. We observe no zero field splitting, which suggests spin-1/2 for these defects.  We use a spin-pair  model\cite{robertson2024universal, gao2024single} to explain experimental observations and also perform first-principles calculations to illuminate the possible structure and properties of these spin defects. Our spin-pair modeling explains the observed Rabi and optically detected magnetic resonance (ODMR) data. We study the coherence times of these spin defects and perform Carr-Purcell-Meiboom-Gill (CPMG) pulse sequence experiments to extend the coherence times by a factor of 20. This study illustrates the potential of $\beta$-GeS$_2$ as a promising 2D material for the fields of quantum sensing.

% We apply a magnetic field perpendicular to the surface of the nanosheets and the microwave magnetic field is along the surface of the nanosheets. This ensures efficient driving of the spin defects under laser excitation. We observe an ODMR contrast of up to 1\% in the $GeS_2$ samples thus prepared, although it varies between 0.35 to 2 \% within the same measurement. 

% \red{Description of band structure, DFT calculations etc, }.

Figure \ref{figure_1}(a) shows the crystal structure of the 2D $\beta$-GeS$_2$ lattice, with the black box highlighting the unit cell in the top view. The bottom part shows the side view of the lattice. 
% The $\beta$-GeS$_2$ flakes used in our experiments are a few tens of nanometers thick. 
The samples are exfoliated from commercially purchased crystals and deposited on Si/SiO$_2$ substrates (supplementary information). We then anneal the samples at 450 $^o$C for 2 hours in air under ambient pressure. The flakes are then transferred to a microwave stripline via a standard polymer based transfer technique, as imaged in Figure \ref{figure_1}(b). We perform confocal scanning, continuous wave ODMR (CW ODMR) and coherent control experiments on the sample. We perform these experiments under 532 nm laser excitation and collect the photoluminescence (PL) emission of the defects. Figure \ref{figure_1}(c) shows the confocal scans of the $\beta$-GeS$_2$ samples under the laser excitation, with the black circles showing the bright spots of interest near the bottom of the image. We select these spots to perform the spin studies as described later in the paper. Figure \ref{figure_1}(d) shows the typical PL spectrum of the spin defects under laser excitation, with a broad spectrum stretching from 550 nm to 900 nm. We also perform Raman spectroscopy (Figure \ref{figure_1}(d), inset) and observe sharp peaks at 360 cm$^{-1}$, which confirms the samples being $\beta$-GeS$_2$ \cite{mitsa2014spectroscopic, yang2019polarization} (Fig S.1, S.4). To perform the CW ODMR experiments we track the PL emission with the microwave (MW) excitation on and off, defining the contrast as $C = \frac{(I_{on} - I_{off})}{I_{off}}$. Figure \ref{figure_1}(e) shows the ODMR contrast at different external magnetic fields. A positive contrast means the PL intensity increases under MW excitation. The peak has a fitted linewidth $\Gamma =$ 187 MHz. The maximum contrast can be as high as 2 \%, though the typical contrast observed is less than 1 \% in our experiments. We observe no zero field splitting for these defects and measure the g-factor of these spin defects by varying the external magnetic field and using carbon-related spin defects in hBN to calibrate the field (Fig S.2). Figure \ref{figure_1}(f) shows the fitted value of the g-factor is $g = 2.000\pm .007$. 
% Rabi oscillation experiments further support this as described in a later section of this paper. 

% Original figure
\begin{figure}
  \includegraphics[scale = .7]{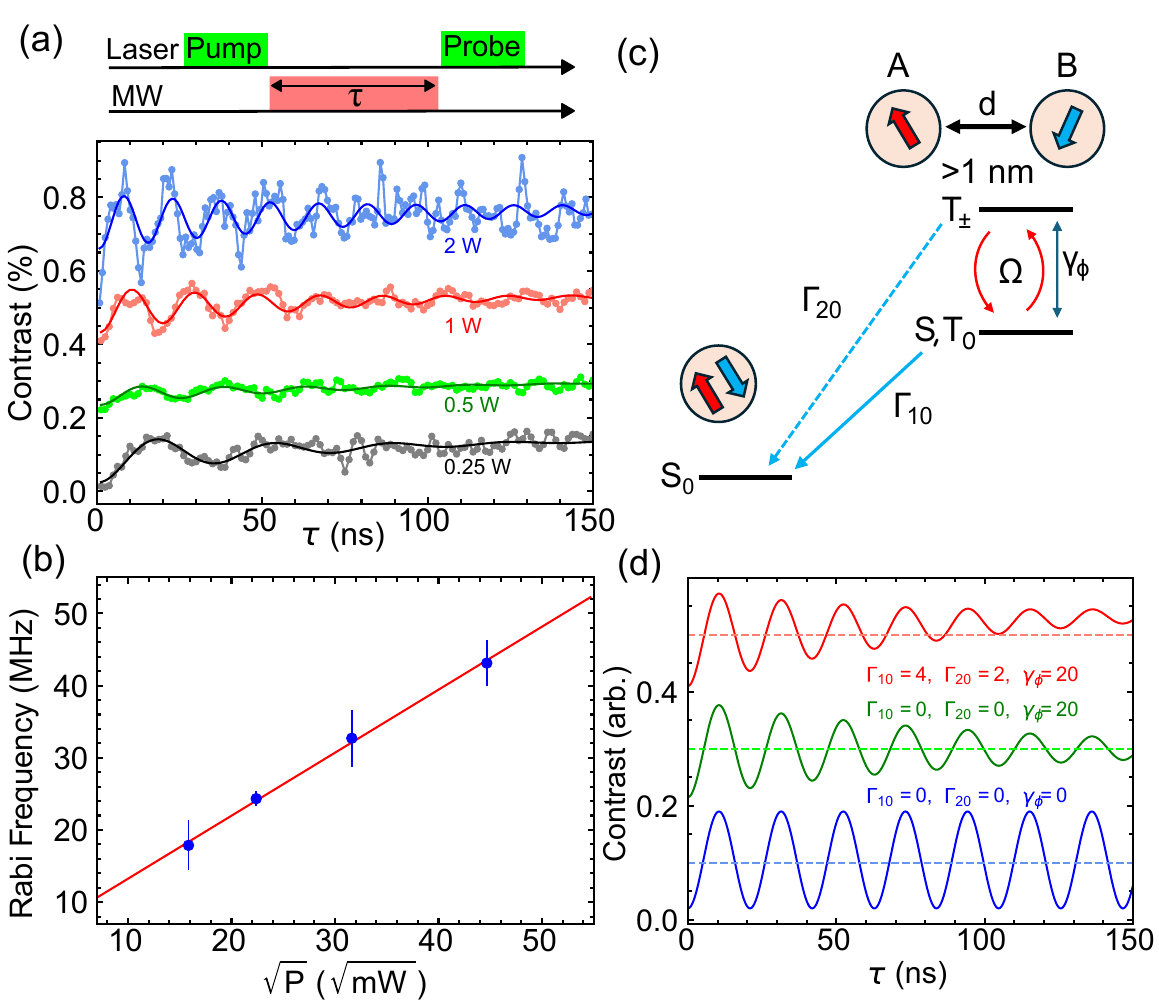}
  \caption{(a) Rabi Oscillations of the spin defects at different MW powers. The solid lines are fits to the spin-pair model predicted oscillations. The top schematic shows the laser and MW pulses used in the experiment. (b) Fitted Rabi frequency vs the square root of the MW power. The solid red line shows a linear fit of the data. (c) The spin pair model consisting of 2 spin defects located a distance of $\gtrsim$ 1 nm from each other. The 3 level system used to model the system is shown at the bottom. The singlet state with both electrons on the same defects is on the left side while the state with the electrons on different defects is on the right. $\Gamma_{10}, \Gamma_{20}$ are the rate constants from $\ket{S,T_0},\ket{T_\pm}$ to $\ket{S_0}$ respectively and $\gamma_\phi$ is the rate between $\ket{S,T_0}$ and $\ket{T_\pm}$. $\Omega$ is the Rabi driving frequency. (d) Simulated ODMR contrast from the spin pair model with different values of the $\Gamma_{10}, \Gamma_{20}$ and $\gamma_\phi$, showing how the model can simulate the various kinds of decay in the Rabi oscillations observed experimentally. All constants have units of $\mu$s$^{-1}$.} 
  \label{figure_2}
\end{figure}

% Trying building figure here itself
% \begin{figure}
% \begin{tcbitemize}[raster equal height=rows,
% raster columns=2, raster halign=center,
% raster every box/.style=blankest]
% \mysubfig{Test}{figures/Figure2/Fig_2_Schematic.pdf}
% \mysubfig{Test}{figures/Figure2/Fig_2_Schematic.pdf}
% \mysubfig{Test}{figures/Figure2/Fig_2_Schematic.pdf}
% \mysubfig{Test}{figures/Figure2/Fig_2_Schematic.pdf}
     % \centering
     % \begin{subfigure}[b]{0.3\textwidth}
     %     \centering
     %     \includegraphics[width=\textwidth]{graph1}
     %     \caption{$y=x$}
     %     \label{fig:y equals x}
     % \end{subfigure}
     % \hfill
     % \begin{subfigure}[b]{0.3\textwidth}
     %     \centering
     %     \includegraphics[width=\textwidth]{graph2}
     %     \caption{$y=3\sin x$}
     %     \label{fig:three sin x}
     % \end{subfigure}
     % \hfill
     % \begin{subfigure}[b]{0.3\textwidth}
     %     \centering
     %     \includegraphics[width=\textwidth]{graph3}
     %     \caption{$y=5/x$}
     %     \label{fig:five over x}
     % \end{subfigure}
     %    \caption{Three simple graphs}
     %    \label{fig:three graphs}
% \end{figure}

In Figure \ref{figure_2}(a) we perform Rabi oscillations at different MW powers of 0.25 - 2 W, with the pulse sequence as shown in the top schematic. A laser pulse of duration 7 $\mu$s polarizes the spin defects into one of the states, and then a resonant microwave drives the Rabi oscillation between the states. Due to different PL emission from each state we see the PL contrast modulation with increasing pulse duration $\tau$. We integrate the collection pulse for a duration of 7 $\mu$s to calculate the PL counts and compare the cases with and without the MW to characterize the ODMR contrast. The Rabi oscillations are seen to lose contrast significantly after ~100 ns and drift upwards asymmetrically, which suggests a metastable configuration similar to
% is characteristic of spin-1/2 defects in
 other systems such as carbon-related defects in hBN \cite{guo2023coherent, scholten2024multi, gao2024single}. 
 
 We therefore adapt the recently proposed spin-pair model for spin-1/2 defects in hBN for this metastable system \cite{gao2024single}. Plotting the fitted Rabi oscillation frequency as a function of the square root of the applied MW power shows a linear dependence as seen in Figure \ref{figure_2}(b). Figure \ref{figure_2}(c) shows the spin pair model used for the defects. Two spin defects A and B are located some distance ($\gtrsim$ 1 nm) apart inside the $\beta$-GeS$_2$ lattice and can undergo charge hopping under laser excitation. When both of them are on the same defect they can form a singlet state. When they are on different defects they form four weakly coupled states. The eigenstates for the weakly coupled spin pair are $T_+ = \ket{\uparrow \uparrow}$, $T_-=\ket{\downarrow \downarrow}$, $T_0 = (\ket{\uparrow \downarrow} + \ket{\downarrow \uparrow})/\sqrt{2}$ and  $S = (\ket{\uparrow \downarrow} - \ket{\downarrow \uparrow})/\sqrt{2}$. For the sake of simplicity this model assumes only 3 states, $\ket {S_0}$, $\ket {S,T_0}$ and $\ket {T_\pm}$. $\ket{S_0}$ consists of the singlet state when both electrons occupy the same defect, $\ket{S,T_0}$ consists of the states $S, T_0$ and $\ket {T_\pm}$ consists of states $T_+$, $T_-$ when the electrons are on different defects. $\Omega$ denotes the Rabi frequency. $\Gamma_{10}$ and $\Gamma_{20}$ are the decay rates from $\ket {S,T_0}, \ket{T_\pm}$ to $\ket{S_0}$ respectively  and $\gamma_\phi$ is the spin-spin relaxation rate between $\ket{S,T_0}$ and $\ket{T_\pm}$. The spin selectivity can be understood in terms of selection rules, wherein the transition between $T_\pm \leftrightarrow S_0$ is forbidden in contrast to the transition $S, T_0 \leftrightarrow S_0$ which is allowed. Using these constants we set up the Lindblad Master Equation to calculate the transitions between the states. The initial state is chosen to be $\ket{S,T_0}$, assuming a polarized initial state. The python toolbox QuTiP\cite{qutip5} is then used to solve for the steady state density matrix from which we extract the populations at different times. We then take a weighted sum of the population to account for the different emissions of each state which yields the simulated ODMR contrast as a function of time. Figure \ref{figure_2}(d) shows simulated ODMR contrast calculated from the solutions of the Lindblad Master equation under different values of the constants $\Gamma_{10}, \Gamma_{20}$ and $\gamma_{\phi}$ and demonstrates how the spin pair model can be used to fit a variety of experimentally observed data. We fit these curves to the Rabi oscillation data in Figure \ref{figure_2}(a) and extract the Rabi frequency along with the other parameters presented in Figure \ref{figure_2}(b).
% \red{describe Rabi Dynamics, spin pair model, weighing the simulated spin dynamics to fit the data, how $\ket{0, 1, 2}$ are made of spin singlet and triplet states}

\begin{figure}
  \includegraphics[scale = .5]{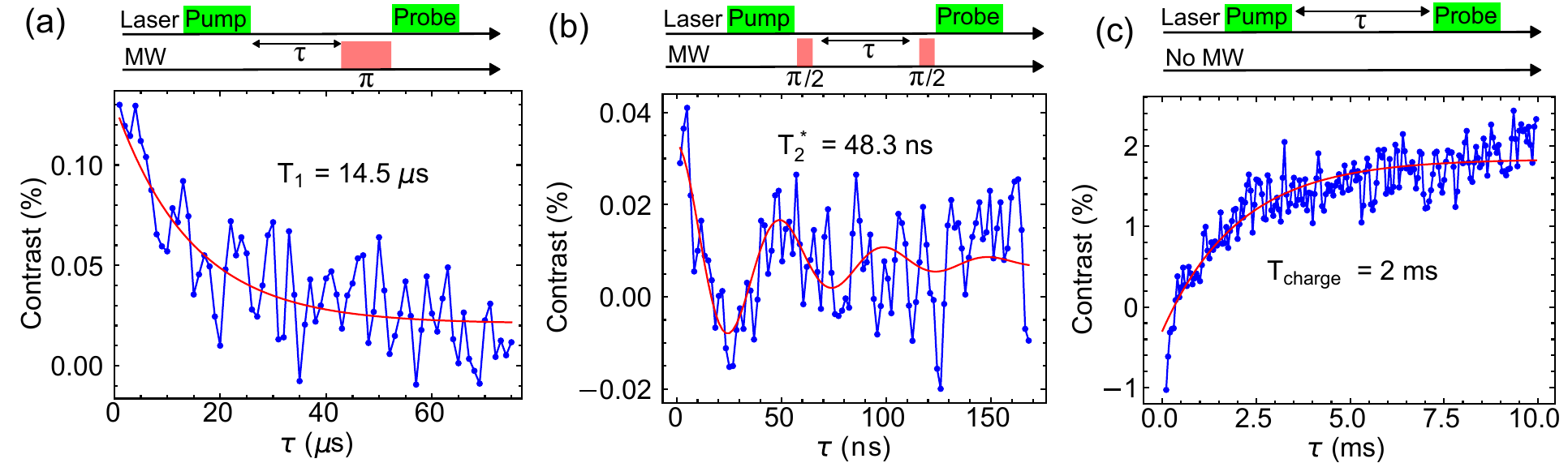}
  \caption{(a) $T_1$ measurements and the pulse sequence schematic used in performing the $T_1$ experiment. The $\pi$ pulse duration is determined from the Rabi experiments as shown in Figure \ref{figure_2}. (b) Ramsey decay for the spin defects. Due to a slight detuning between the resonant frequency and the drive, some oscillations are visible. (c) Experimentally observed charged state dynamics for the spin defects. The top schematic shows the laser pulse sequence used in the experiment. Note that this measurement is performed without any MW applied, and hence only measures evolution of the polarized state with time.
  }
  \label{figure_3}
\end{figure}

We then further probe the spin coherence properties of the spin defects by looking at the spin relaxation time $T_1$.  We first measure the $T_1$ of the spin defects via the pulse sequence shown on top of Figure \ref{figure_3}(a). After polarization by a laser pulse we vary the dark time between pulses and then flip the state via a $\pi$ pulse before reading it out by the next laser pulse. To measure the contrast we compare the case with and without the MW $\pi$ pulse. After fitting to a simple exponential decay curve as $a e^{-t/T_1}+c$, we find that the $T_1$ is 14.5 $\pm$ 4.3 $\mu s$. One possible explanation of the short $T_1$ times can be a large density of emitters close to each other, as can be seen in \ref{figure_1} (c). It would also explain the small ODMR contrast observed in these samples\cite{stern2022room}. We also measure the Ramsey decay via the pulse sequence as shown on top of Figure \ref{figure_3}(b). A $\pi/2$ pulse initializes the state into a superposition of $\ket0$ and $\ket1$ which is then allowed to evolve freely for a time $\tau$. It is then flipped out of superposition by another $\pi/2$ pulse and read out. This contrast as a function of time is fitted to an oscillating exponential decay as $a e^{-t/T_2^*} sin(\omega t + \phi_0)+ c$ which yields the time $T_2^*$ = 48.3 $\pm$ 18 ns. We then probe the PL settling recovery dynamics by means of the pulse sequence shown in top of Figure \ref{figure_3}(c). After polarization the system is allowed to evolve for a time $\tau$ after which it is read out by another laser pulse. This is performed without a MW pulse, to look at only the charge state dynamics and evolution. To calculate the contrast we compare the PL at the end of the polarization pulse and the beginning of the readout pulse. %This shows a continuous increase in the measured contrast which is then fit to a streched exponential curve. The fitted parameters of $\tau = $773 $ \mu s$ and stretching $\beta = 0.227$ suggest a range of exponential decay times across multiple orders of time.
This shows a continuous increase in the measured contrast which is then fit to an exponential curve as $a*e^{-t/T_{charge}}+c$, yielding a timescale $T_{charge}$ = 2 $\pm$ .23 ms.
% 2.03 $\pm$ .23 ms.
This suggests that the charge dynamics of the triplet to singlet transition span a large temporal scale. This further strengthens the claim that this is an ensemble of weakly-coupled spin pair systems. The longer time decays are not expected to affect the ODMR and Rabi measurements due to the short collection times during initialization and collection and the short evolution times in those measurements. 
% \red{Explain why $T_1$ is so short, talk about shelving dynamics, explanation about pl recovery being due to spin pair recombination}.

\begin{figure}
  \includegraphics[scale = .45]{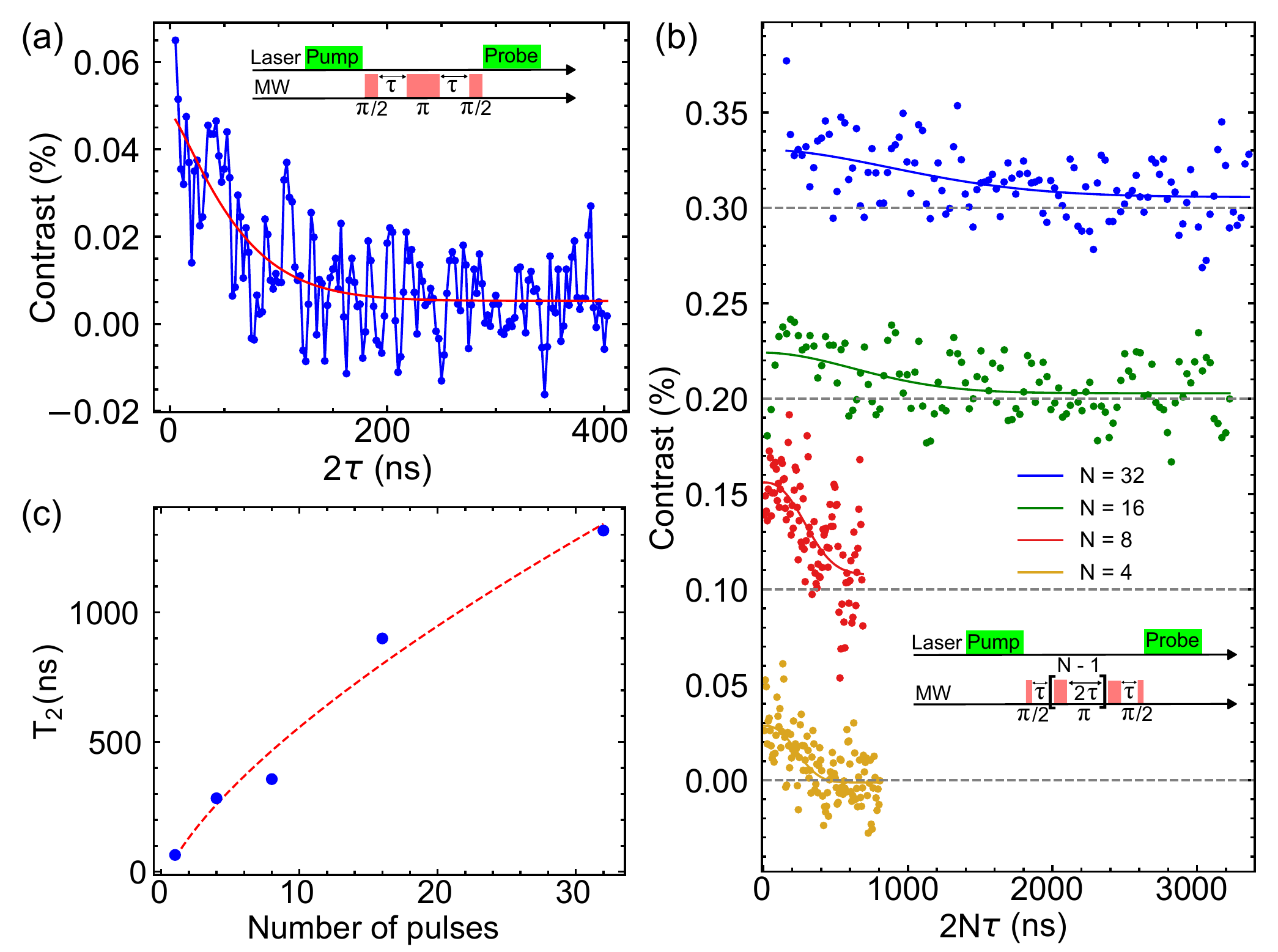}
  \caption{(a) Measurement of the coherence time $T_2$ of the spin defects in $\beta$-GeS$_2$ via the Hahn Echo pulse sequence as shown in the schematic. The red line shows a stretched exponential fit. (b) Extending the $T_2$ using the CPMG pulse sequence for different number N of $\pi$ pulses. The schematic shows the pulse sequence used in the experiment for each N. (c) The extended $T_2$ plotted as a function of number of pulses in the sequence and fitted to a power law $\propto a N^\gamma$ with $\gamma$ = 0.69.}
  \label{figure_4}
\end{figure}

Dynamical decoupling techniques have been used for extending the coherence times of spin qubits in hBN \cite{rizzato2023extending} . Due to sample impurities and high density of emitters in $\beta$-GeS$_2$, it is not surprising that the coherence times are short, as is seen in the $T_1$ measurements as well \cite{haykal2022decoherence, gong2024isotope}. In order to probe the limits of spin coherence in $\beta$-GeS$_2$ we apply these techniques and measure the $T_2$. Figure \ref{figure_4}(a) shows the measured coherence time $T_2$ for the $\beta$-GeS$_2$ spin defects measured via the Hahn Echo pulse sequence. The inset shows the pulse sequence. A laser pulse of duration 7 $\mu s$ polarizes the spins, which are then driven to a superposition state by a $\pi/2$ pulse. After a time $\tau$, a $\pi$ pulse flips the spin followed by another $\pi/2$ pulse after a time $\tau$. The spins are then read out by a laser pulse. The measured PL after the pulse sequence for $\tau$ varying from 5 to 400 ns shows a decay which is then fit to an exponential function  to extract the $T_2$ time of the spins. This $T_2$ is measured to be 64.5 $\pm$ 12 ns. This is not unexpected since we observe an ensemble of spin defects which occur in clusters and can have significant interactions with each other and with surrounding impurity nuclei, reducing the spin coherence times. We then attempt to decouple the spins from external magnetic noise by the CPMG pulse sequence\cite{meiboom1958modified, rizzato2023extending}. A laser pulse initializes the spins which are then driven to superposition by a $\pi/2$ pulse. This is then followed by N separate $\pi$ pulses, separated by a time $2\tau$ and then another $\pi/2$ pulse after a delay $\tau$. The spins are then read out and the measured contrast plotted as a function of the total time duration 2N$\tau$. Figure \ref{figure_4}(b) shows extending of the coherence time using the CPMG sequence with N varying from 4 to 32. We fit the decay to a stretched exponential of the form $e^{-(t/T)^\beta}$ and find that the $T_2$ coherence time can be extended from 64.5 ns as measured by the Hahn Echo pulse sequence to 1.315 $\mu s$ in the case of CPMG-32 pulses. Figure \ref{figure_4}(c) shows the extended $T_2$ times plotted as a function of number of pulses in the CPMG pulse sequence and fitted to a power law function $T_2 \propto a N^\gamma$ where $\gamma$ is found to be 0.691, which matches closely with the theoretical value of 2/3 for a Lorentzian spin bath\cite{rizzato2023extending}. This suggests that with better material growth techniques it might be possible to observe even longer coherence times \cite{ bar2012suppression, pham2012enhanced, haykal2022decoherence, rizzato2023extending}. We observe a factor of 20 increase in the coherence time under the CPMG sequence, and with more pulses and tighter pulse sequence control it is possible for the $T_2$ to approach $T_1$ times.
% \red{ Talk about the NatCom paper HahnEcho times, justification for CPMG pulses etc}

\begin{figure}
  \includegraphics[scale = .5]{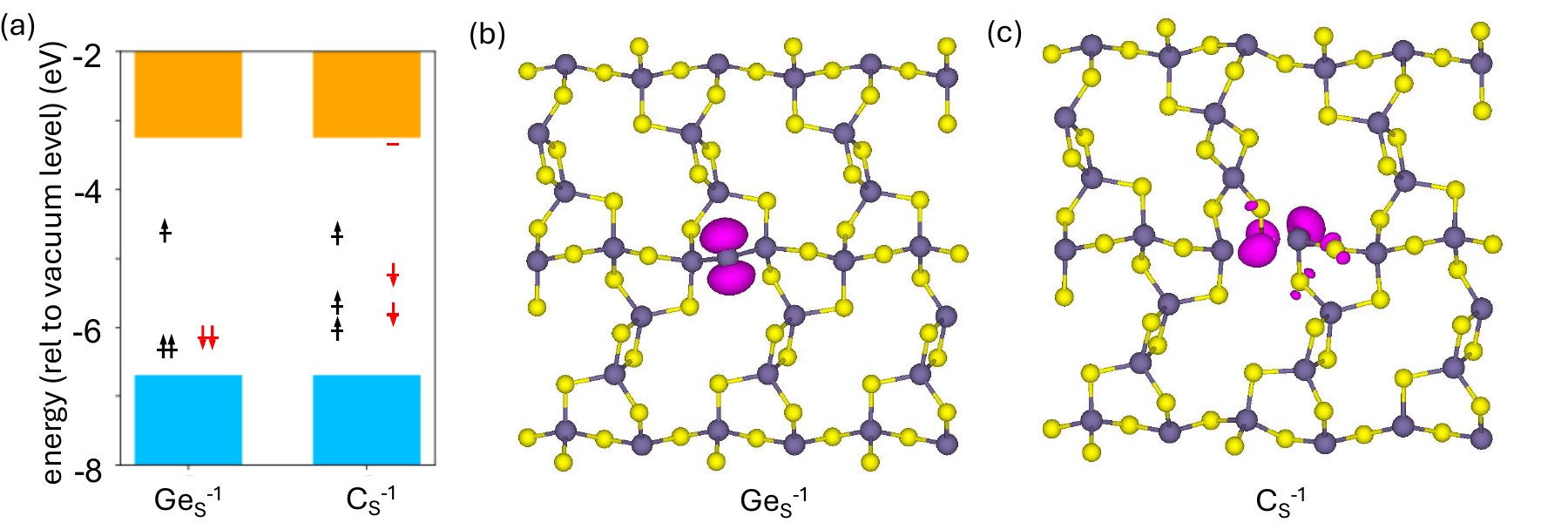}
  \caption{(a) Energy diagram showing the energy levels of the two possible defects in $\beta$-GeS$_2$ in the bandgap. (b), (c) Calculated spin density for the Ge$_S^{-1}$ and C$_S^{-1}$ defects in the lattice respectively.}
  \label{figure_6}
\end{figure}

To shed light on the possible structure of these spin defects in $\beta$-GeS$_2$, we performed first-principles calculations on the $\beta$-GeS$_2$ in the 2×1×1 supercell. Among the native defects (vacancies and antisite defects), substitution defects (with external species of carbon, nitrogen, and oxygen), and defect complex V$_{Ge}$C$_S$, we identified the negatively charged defects Ge$_S^{-1}$ and C$_S^{-1}$ as potential spin defect candidates with unpaired spin ground states. The calculated defect formation energies (see supporting information) are 4.14 and 2.23 eV, respectively\cite{van2004first, leem2024optically}. Figure \ref{figure_6}(a) shows the calculated energy levels of these defects with respect to the vacuum level, where the arrows represent electrons in the spin-up (black) and spin-down (red) channels. Both Ge$_S^{-1}$ and C$_S^{-1}$ have an unpaired electron, which is further validated by the spin density. The spin density, defined as the difference of density of the spin-up and spin-down channels, are shown in Figure \ref{figure_6}(b),(c), suggesting well-localized spin defect states. For Ge$_S^{-1}$, the unpaired spin is localized on the Ge atom and is primarily of p-orbital character, while for C$_S^{-1}$, the spin is mainly localized on the C atom and nearby Ge atom. To satisfy the spin pair model, there can be another dark spin nearby consisting of these defects or some other defects. The antisite defect Ge$_S^{-1}$ introduces multiple in-gap states, which can facilitate defect-to-band optical transitions. In the case of C$_S^{-1}$, the transitions between the defect states in the spin-down channel are plausible.

% \red{ How to improve: material growth, less impurity, low temperature, read more papers}

In conclusion, we have observed optically active spin defects in a new 2D material platform $\beta$-GeS$_2$ which may potentially host long lived electron spin coherence at room temperature. 
% While the currently observed spin properties are not state of the art, with advances in materials growth and purification it should be possible to improve these numbers by a few orders of magnitude. 
We measure $T_1$ of 14.5 $\mu$s and extend the $T_2$  by a factor of 20 to 1.3 $\mu$s. While short, there is considerable scope of improvements in these numbers with advances in crystal growth.
% , isotopic purification such as in diamond\cite{balasubramanian2009ultralong} and hBN\cite{clua2023isotopic} and defect creation techniques. 
The next step would be to study these defects at low temperature and see the effects of thermal noise on the electron spin properties. Also, in line with the carbon-related spin defects in hBN, observation of single optically active spin defects in $\beta$-GeS$_2$ would shed further light on the nature of the defects and enable study of their properties in more detail. The low nuclear spin density and the potential for achieving a nuclear spin-free composition makes this material promising for a wide variety of quantum technologies where minimizing environmental interference into the dynamics of the electron spin qubits is essential. 
% With these advances, this material platform could prove to be a very effective tool for quantum sensing, computing and simulation.
% In conclusion, we have demonstrated GeS$_2$ as a van der Waals material hosting spin defects. 
 % Being a 2D material, GeS$_2$ is also potentially easily integrable with the semiconductor processes, which would speed up its integration into the next generation of quantum technologies. 
  Given that the theoretically calculated upper limit of coherence times in GeS$_2$ is 4.5 ms\cite{kanai2022generalized, sajid2022spin, fabian2022quantum}, this system is highly interesting for the field of spin-defect color centers.
\begin{acknowledgement}

The authors thank Zhiyan Xie and Nithin Abraham for fruitful discussions and suggestions. 
This work was mainly supported by the U.S. Department of Energy, Office of Science, Office of Basic Energy Sciences through Quantum Photonics Integrated Design Center (QuPIDC) Energy Frontier Research Center under award number DE-SC0025620. T.L. also acknowledges the support by the Gordon and Betty Moore Foundation, grant DOI 10.37807/gbmf12259. Z.F. and Q.Y. acknowledge the support by the National Science Foundation under Grant No. DMR-2314050.

\end{acknowledgement}

%%%%%%%%%%%%%%%%%%%%%%%%%%%%%%%%%%%%%%%%%%%%%%%%%%%%%%%%%%%%%%%%%%%%% 
%% The same is true for Supporting Information, which should use the
%% suppinfo environment.
%%%%%%%%%%%%%%%%%%%%%%%%%%%%%%%%%%%%%%%%%%%%%%%%%%%%%%%%%%%%%%%%%%%%%
\begin{suppinfo}

The Supporting Information is available free of charge.

\end{suppinfo}

\section{Note}
There is a relevant preprint\cite{liu2024experimental} reporting spin defects in $\beta$-GeS$_2$.

%%%%%%%%%%%%%%%%%%%%%%%%%%%%%%%%%%%%%%%%%%%%%%%%%%%%%%%%%%%%%%%%%%%%%
%% The appropriate \bibliography command should be placed here.
%% Notice that the class file automatically sets \bibliographystyle
%% and also names the section correctly.
%%%%%%%%%%%%%%%%%%%%%%%%%%%%%%%%%%%%%%%%%%%%%%%%%%%%%%%%%%%%%%%%%%%%%
\bibliography{achemso-demo}

\end{document}